\documentclass[preprint,showpacs,showkeys,preprintnumbers,amsmath,amssymb]{revtex4}

\usepackage{graphicx}
\usepackage{dcolumn}
\usepackage{bm}

\newcommand{\bb}{\backslash}
\newcommand{\sign}{\mathrm{sign}}

\begin{document}

\preprint{APS/123-QED}

\title{APPLICATIONS OF CORRELATION INEQUALITIES TO LOW DENSITY GRAPHICAL CODES}

\author{Nicolas Macris}
 \email{nicolas.macris@epfl.ch}
\affiliation{
Laboratoire de Th\'eorie des Communications\\
Ecole Polytechnique F\'ed\'erale de Lausanne\\
Station 14 - LTHC -EPFL\\
CH - 1015 Lausanne, Switzerland}

\date{September 30, 2005}

\begin{abstract}
This contribution is based on the contents of a talk delivered at the Next-SigmaPhi conference held in Crete in August 2005.
It is adressed to an audience of physicists with diverse horizons and does not assume any background
in communications theory. 
Capacity approaching error correcting codes for channel communication known as 
Low Density Parity Check (LDPC) codes have attracted considerable attention
from coding theorists in the last decade. 
Surprisingly strong connections with the theory of diluted
 spin glasses have been discovered. In this work we elucidate one new connection, namely that a class of correlation inequalities
valid for gaussian spin glasses can be applied to the theoretical analysis of LDPC codes. This allows for a rigorous comparison between the so 
called (optimal) maximum a posteriori and
the computationaly efficient belief propagation decoders. The main ideas of the proofs are explained and we refer to recent works for the more lengthy technical details. 
\end{abstract}

\pacs{05.20.-y,  89.70.+c, 02.90.+p}
                             
\keywords{Spin glass, error correcting code, belief propagation, correlation inequality }
\maketitle

\begin{section}{Codes for communication through noisy channels}

We consider a (simplified) communication system with three basic building blocks: the 
{\it encoder}, the {\it channel} and the {\it decoder}.

\noindent{\it Encoder.}
Suppose that messages to be sent are labelled $\{1,...,M\}$ and that $M=2^K$. The messages can be represented by binary strings of length $K$, so that if a message is sent $K$ {\it information bits} are transmitted.
 Because of channel imperfections these binary strings are encoded before they are fed into the channel.
In general the encoder is a map $\mathbb{F}_2^K \rightarrow \mathbb{F}_2^N$, with $\mathbb{F}_2=\{0,1\}$ and $N>K$. 
 So the codebook consists of $2^K$ code words  that are binary strings of length $N$, $(x_1,...,x_N)={\bf x}$. In order to 
 send $K$ information bits we make $N$ uses of the channel: one says that the rate of transmission is 
$R=\frac{K}{N}$. 

\noindent{\it Channel.}
We take a 
discrete (binary input) memoryless channel with general output alphabet (for example $\mathbb{F}_2$ or $\mathbb{R}$). Given a sent codeword $(x_1,...,x_N)$ the received word
is $(y_1,...,y_N)={\bf y}$ with probability
$
p_{{\bf Y}\vert {\bf X}}({\bf y}\vert {\bf x}) = \prod_{i=1}^N p_{Y\vert X}( y_i\vert x_i)
$
In this context the choice of the transition probability $p_{Y\vert X}$ specifies the model for the channel and is supposed to be known to the sender and the receiver.

\noindent{\it Decoder.}
Given that ${\bf x}^{in}$ is sent, the receiver possesses a deformed version ${\bf y}$ (the channel
observations or the channel output)  and his task is to find estimates  $D({\bf y})$ so that the bit probability 
of error $P_{error}((D({\bf y}))_i\neq x_i^{in})$ is as small as possible. One can show that the best decoder (the one which gives the smallest probability of error) 
is given by the Maximum a Posteriori
 (MAP) estimator
$
 (\hat x_i)_{MAP}= \mathrm{argmax}_{x_i} p_{X\vert {\bf Y}}(x_i\vert {\bf y})
$
Unfortunately this cannot be computed efficiently and other suboptimal estimators must be considered.
Of course it is important to compare their relationship and performance to the MAP estimator. This 
problem is adressed here for LDPC codes and the suboptimal estimator given by Belief Propagation (BP).

Shannon's  
{\it noisy channel coding theorem} asserts that one can communicate reliably as long as the rate $R$ is smaller than 
the {\it channel capacity} $C=\mathrm{max_{p_{\bf X}}} I(X;Y)$. In this formula $I(X;Y)$ is the {\it mutual information} between random variables $X$ and $Y$ 
which can be interpreted as the information gained about $X$ given that $Y$ is observed. The maximization over the prior distribution of 
the codewords $p_{X}$ corresponds to finding the best possible codebook. In formulas, 
$I(X;Y)=H({X}) - H({X}\vert  Y) = H({Y}) - H({Y}\vert  X)$,
where the Shannon entropy of $X$ is $H({X})=-\sum_{x}p_{ X}({x})\ln p_{ X}({ x})$
and the conditional entropy 
$H( X\vert  Y)=-\sum_{ x,y}p_{ Y}({ y})p_{{ X}\vert { Y}}({ x}\vert{ y})
\ln p_{{ X}\vert { Y}}({ x}\vert{ y})$
and similarly for $X$ and $Y$ exchanged.
All marginals are computed from
$
p_{{\bf X},{\bf Y}}({\bf x},{\bf y})=p_{\bf X}({\bf x})p_{{\bf Y}\vert {\bf X}}({\bf y}\vert {\bf x})
$.
Thus $C$ is a functional of the channel transition probability.
Moreover there is no way to communicate reliably when $R>C$. 

More precisely, let $R\leq C-\epsilon$ where $\epsilon>0$ is as small as we wish. There exists an $N_0(\epsilon)$ such that for each $N>N_0(\epsilon)$ we can find encoding
and  decoding maps (the optimal decoder does the job) such that  $P_{error}<\epsilon$.
Conversely, if $R\geq C-\epsilon$ for any $N$ and any encoding map $P_{error}\geq p_0>0$
for some $p_0$ independent of $N$.

For our purpose it is more convenient to fix a desired rate $R$ once for all and translate the inequality 
$R<C$ as a condition on the channel noise $n<n_{sh}$ where $n_{sh}$ is a (channel dependent) function of $R$. This means we can reliably transmit at rate 
$R$ as long as the channel noise is lower than the Shannon threshold $n_{sh}$.

Shannon's theorem is not constructive in the sense that it garantees the existence of an encoder in an ensemble of random codes,
but does not allow to construct "good"
(capacity approaching and computationaly efficient) encoders
and decoders. One of the main themes of information and coding theory for the last fifty years has 
been to precisely define and address such questions. A fruitful idea is to restrict the encoder maps
to the class of linear error correcting codes. Remarkably Shannon's theorem is still true if one restricts to the class of linear encoders and there is no loss in capacity.
For more details we refer the reader to \cite{Ash}.

For us a linear code is a vector subspace of 
$\mathbb{F}_2^N$ of dimension $K<N$. The subspace can be defined as the kernel of a parity check $M\times N$ matrix $H$ with $N-M=K$. In other words the set of code words satisfy $M$ constraints (so called parity checks)
\begin{equation}\label{paritycheck}
\sum_{k=1}^{N} H_{lk} x_k=0\,\, mod 2,\,\, l=1,...,M,\,\, H_{lk}=0,1
\end{equation}
Note that the rate of the code is $R=\frac{K}{N}=1-\frac{M}{N}$. A very useful graphical representation of a linear code is in terms of the Tanner graph (or factor graph). This is a bipartite graph
with variable nodes $i\in \{1,...,N\}$, check nodes $A\in \{1,...M\}$, and edges connecting variable  and check nodes. We say that a variable node $i$ "belongs" 
to a check node $A$, $i\in A$, if and only if
it appears in the parity check equation labeled by $A$. In this case an edge connects $i$ and $A$ (see figure 1). 
\begin{figure}
\begin{center}
\includegraphics[ width=0.5in,height=2.5in , angle=90]{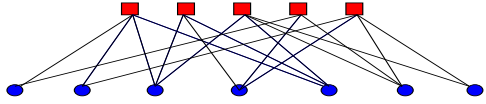}
\caption{\label{fig:1} A Tanner graph. Check nodes on the top row constrain the bits attached to variable nodes on the bottom row}
\end{center}
\end{figure}
Low Density Parity Check (LDPC) codes are a special class of linear codes with sparse Tanner graphs:
the degrees (or coordination number)  of check and variable nodes are of $O(1)$ with respect to $N$.
For such codes there is still a threshold phenomenon as in Shannon's theorem however 
in general the maximal rate at which error free communication is possible is below Shannon's capacity. 
On the other hand suboptimal but computationaly efficient 
decoding algorithms exist.   

\end{section}

\begin{section}{Low Density Parity check codes as diluted spin glasses}

The close connection of the above formalism to random spin systems was first noticed by 
Sourlas \cite{Sourlas}. While this connection is quite general and not limited to binary alphabets, memoryless channels and linear codes, 
here we rephrase it in the case of low density parity check codes. If code word bits are represented by spins through the mapping
 $s_i=(-1)^{x_i}$, the parity check equations \eqref{paritycheck} become  
\begin{equation}
\frac{1}{2}(1+s_A)=1,\qquad 
s_A=\prod_{ i\in A}s_{i},\qquad A=1,...,M
\end{equation}
The a posteriori probability distribution used in MAP decoding is nothing else than the Gibbs measure
of a spin system where the spins are attached to variable nodes while check nodes are a convenient way to represent their many-body interactions. 
By  Bayes rule
\begin{equation}
p_{{\bf X}\vert {\bf Y}}({\bf x}\vert {\bf y})= \frac{1_{\cal C}({\bf x})\prod_{i=1}^N p_{Y\vert X}(y_i\vert x_i)}{\sum_{{\bf x}}
1_{\cal C}({\bf x})\prod_{i=1}^N p(y_i\vert x_i)}
\end{equation}
This is a Gibbs Measure $\langle-\rangle_{\cal C}=
\frac{e^{-H_{\cal C}}}{Z_{\cal C}}$ with hamiltonian
\begin{equation}
H_{\cal C}=-\sum_{A\in{\cal C}} {J_A} (s_A-1)- \sum_{i=1}^n h_i s_i,\qquad s_{A}=\prod_{i\in A}s_i
\end{equation}
where $J_A=+\infty$ and $h_i=\frac{1}{2}\ln\frac{p(y_i\vert 0)}{p(y_i\vert 1)}$. The channel observations enter through a quenched random magnetic field $h_i$
whose distribution is induced by the distribution of channel observations. It can be shown that for symmetric channels (these satisfy $p(y\vert x)=
p(-y\vert -x)$) there is no loss in generality to assume that the input word is $(x_1^{in}=0,...,x_N^{in}=0)$, so that the distribution of channel observations is $\prod_{i=1}^{N}p(y_i\vert 0)$.
Another source of quenched randomness is given by the Tanner graph (defining the coupling constants $J_A$) which is taken from an ensemble of random graphs.
Since our results are independent of the choice of this ensemble we do not discuss their construction in detail. Let us point out that the performance of a particular coding scheme
depends on the choice of the ensemble.
The expectation value with respect to the channel observations and the graphs are denoted $\mathbb{E}_{\mathcal{C}, h}$.
 
The MAP decoding rule becomes
\begin{equation}
(\hat s_i)_{MAP}= sign \langle s_i \rangle_{\cal C}
\end{equation}
and the average bit probability of error for the optimal decoder is basically the overlap of 
$(\hat s_1,...,\hat s_N)$ with the fully ferromagnetic configuration $(1,...,1)$ (or the sent codeword)
\begin{equation}\label{proberror}
P_{error} = \frac{1}{N}\sum_{i=1}^N\mathbb{E}_{\mathcal{C}, h}[1 - sign \langle s_i \rangle_{\cal C}]
\end{equation}
The replica or cavity methods can be applied to the calculation of such quantities and show that a phase transition occurs \cite{Montanari1}. 
Namely there is a threshold $n_{MAP}$ such that for $n<n_{MAP}$ the probability of error goes to zero in the thermodynamic limit (ferromagnetic phase), 
while for $n>n_{MAP}$ the probability of error is bounded away from zero.
Sparse graphs are localy tree like in the sense that the typical size of loops is $O(N)$ and have no boundary. 
Hence it is reasonable to expect that mean field approaches such as the replica or cavity methods yield exact results. This is for the moment unproven although some progress in this direction has been made
by the use of interpolation methods \cite{Guerra}, \cite{Montanari2}. As explained below our use of correlation inequalities yields
closely related results.

\end{section}

\begin{section}{Efficient decoding} 

Although one can optimize the degrees of the Tanner graphs in order that $n_{MAP}$ approaches 
$n_{sh}$, MAP decoding is computationally too expensive. However one can take advantage of the fact that low density graphs are 
 localy tree like (see figure 2). Consider a specified root 
 node $o$ and its neighborhood of depth $d$.
As long as $d=O(1)$ with respect to $N$ this neighborhood is a tree with high probability. 
Thus one can expect that a good approximation is obtained by neglecting the loops and solving
for the magnetization of the spin system on a tree.
\begin{figure}
\begin{center}
\includegraphics[width=3in, height=1in]{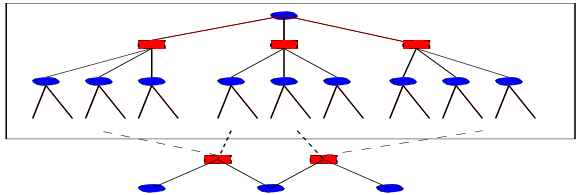}
\caption{\label{fig:2} Tree like neighborhood $T_o$ of an arbitary root node $o$. The loops are of size $O(N)$ with high probability}
\end{center}
\end{figure}
The sign of the magnetization on the tree defines the {\it Belief Propagation} (BP) estimate
\begin{equation}
(\hat s_o)_{BP}= \sign\tanh(h_o +\sum_{C\in o} u_{C\to o}^{(d-1)})
\end{equation}
In this formula the fields $u_{C\to i}$ are computed from the iterative procedure
\begin{equation}
u_{C\to i}^{(t+1)}=\tanh^{-1}\prod_{j\in C\bb i} \tanh h_{j\to C}^{(t)},\qquad
h_{i\to A}^{(t+2)}= h_i+\sum_{C\in V(i)\bb A} u_{C\to i}^{(t+1)}
\end{equation}
with the initial conditions $h_{i\to C}^{(0)}=h_i$.

The belief propagation decoding algorithm is an iteration based on these exchanges of {\it messages} $u_{C\to i}$ from checks to variables and {\it messages}
$h_{j\to C}$ from variables to checks. It is applied to the full Tanner graph and despite the presence of loops it converges and succesfully decodes for $n<n_{BP}$.
The relationship between the various thresholds is $n_{BP}<n_{MAP}<n_{sh}$.
It should be clear that this algorithm is closely related to 
the
cavity equations of spin glass theory. 

One of the main problems in the theory of LDPC codes is 
to optimize the codes so that the various thresholds come as close as posible to $n_{sh}$. 
A more basic problem is to compare the error probabilities given by the BP and MAP decoders.
While this is difficult in general we show below how these decoders can be compared for closely related
quantities - {\it the generalized EXIT curves} - through the use of correlation inequalities.

\end{section}

\begin{section}{Correlation inequalities}

Here we restrict ourselves to the case of 
the binary input additive white gaussian noise channel (BIAWGNC) where the results are more transparent. Mathematicaly the channel is defined as
$y_i= x_i + W_i$, $W_i$  i.i.d ${\cal N}(0, n)$.
Then the log-likelihood ratio (or magnetic field)
$h_i=\frac{1}{2}\ln\frac{p(y_i\vert0)}{p(y_i\vert 1)}$ has a {\it gaussian} distribution with equal 
mean and variance $\mathbb{E}_h[h_i]=\mathbb{V}_h[h_i]=n^{-1/2}$.
We soften the  parity check constraints from $J_A=+\infty$ to independent gaussian random variables with equal mean and variance 
$\mathbb{E}_J[J_A]=\mathbb{V}_J[J_A]=t_A$. The case of hard constraints
(the parity checks) is recovered by making $t_A\to+\infty$.
With solft random constraints 
the hamiltonian is  a {\it gaussian spin glass} with Nishimori gauge symmetry. Contucci, Graffi and Nishimori proved for such systems the following set of inequalities hold
\cite{Contucci}
\begin{equation}
\mathbb{E}_J[\langle s_X\rangle]\geq 0,\qquad \frac{\partial}{\partial t_Y} \mathbb{E}_J[\langle s_X\rangle]\geq 0,\qquad \mathrm{any}\qquad X,Y\subset\{1,...,N\}
\end{equation}
The reader will recognize the close similarity to the famous  Griffith-Kelly-Sherman correlation inequalities valid for fully ferromagnetic systems.

This inequality can be applied to compare the magnetization on the initial Tanner graph and on a tree graph. In the coding context this allows a comparison 
between MAP and BP decoders.
Consider the Gibbs measure defined by the gaussian spin glass hamiltonian with some set of variances $t_A$, $A=1,...,M$. 
The neighborhood $T_o$ of $o$ (see figure 3) is a tree with probability $(1 -O(\frac{k^d}{N}))$ where $k$ is a constant related to the maximal degree of the nodes.
 The second correlation inequality implies that, if for
 the checks outside of $T_o$ we decrease
$t_A$ to zero, the average magnetization of site $o$ decreases. This inequality is preserved if we increase $t_A$ to infinity for the checks inside $T_o$.
In other words
\begin{equation}\label{correlation}
\mathbb{E}_{{\cal C},h}[\langle s_o\rangle_{\cal C}]\geq \mathbb{E}_{{\cal C},h}[\langle s_o\rangle_{ T_o}\vert T_o ~is ~ a~tree]Pr(T_o~is~a~tree)
\end{equation}
The right hand side should also incude a contribution coming from the probability that $T_o$ is not a tree but by the first correlation inequality it is positive so that we can omit it. We refer to this
procedure as the "check erasing" (see figure 3 for a pictorial illustration of check erasing).
\begin{figure}
\begin{center}
\includegraphics[width=2in, height=1in]{inequgraph1} \,\,
\includegraphics[width=0.25in]{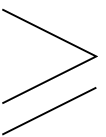}\,\,
\includegraphics[width=2in, height=1in]{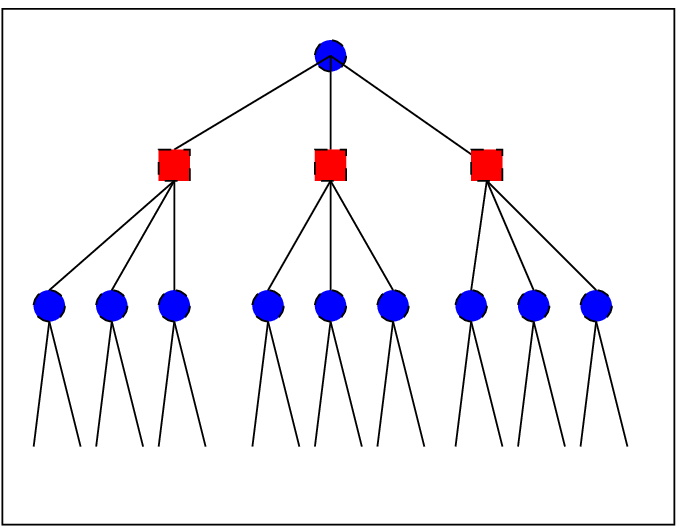}
\caption{\label{fig:3}A pictorial representation of the check erasing inequality}
\end{center}
\end{figure}
On the tree graph the statistical mechanical sums can be performed exactly and yield in a natural way the Belief Propagation algorithm of the previous section. So
\begin{equation}
\mathbb{E}_{{\cal C},h}[\langle s_o\rangle_{\cal C}]\geq (\hat s_o)_{BP}(1 -O(\frac{k^d}{N}))
\end{equation}
Finaly one can take the thermodynamic limit $N\to+\infty$ and then the limit $d\to+\infty$. While on the right hand side these limits can be shown to exist,
the existence of the thermodynamic limit for the left hand side is an open problem. Thus we realy take the $\liminf_{N\to\infty}$.

\end{section}

\begin{section}{Generalized EXIT curves}

The probability of error \eqref{proberror} is technically cumbersome to handle. Another quantity called in coding theory the "extrinsic information transfer"
is more convenient to study. It yields the same thresholds as the error probability and from the satistical mechanical
perspective it is much more natural as will become clear below. Here we define the generalized EXIT curve associated to MAP decoding as \cite{Urbankebook}
\begin{equation}
g_{MAP}(n)=\liminf_{N\to\infty}\frac{1}{N}\frac{d}{dn} \mathbb{E}_{\cal C}[H(X_1,...,X_N\vert Y_1,...,Y_N)] 
\end{equation}
The conditional entropy of the a posteriori distribution is nothing else than the average entropy of the Gibbs distribution
for the spin glass. It should not come as a surprise that this can be related to the free energy
\begin{equation}\label{free}
\mathbb{E}_{\cal C}[H({\bf X}\vert {\bf Y})]= \mathbb{E}_{{\cal C},h}[\ln Z_{\cal C}] - \sum_{i=1}^N \mathbb{E}_{{\cal C},h}[h_i\langle s_i\rangle_C]
\end{equation}
In the case of a BIAWGNC the derivative with respect to the noise has a simple relation to the magnetization. This is not obvious a priori because the channel noise
does not enter like an external field and for more general channels the corresponding relation is more complicated. 
The derivation of \eqref{free}, \eqref{identity} is too lengthy to show here but let us note that the main
point is to use Nishimori identities \cite{Nishimori}
\begin{equation}\label{identity}
g_{MAP}(n)=\liminf_{N\to+\infty}\frac{1}{2n^3N}\sum_{i=1}^{N}\mathbb{E}_{{\cal C},h}[1-\langle s_i\rangle_{\cal C}]
= \frac{1}{2n^3}\mathbb{E}_{{\cal C},h}[1-\langle s_o\rangle_{\cal C}], \qquad \mathrm{any~} o
\end{equation}
The following lemma shows that $g_{MAP}(n)$ and $P_{error}$ have the same threshold.

\vskip 0.25cm

\noindent{\bf Lemma. }{\it 
Assume communication through a BIAWGNC with noise $n$ and an ensemble of linear codes. We have that
$g_{MAP}(n)=0$ if and only if $P_{e}=\lim_{N\to +\infty}P_{error}=0$.}

\vskip 0.25cm

To show that $g_{MAP}(n)=0$ implies $P_{e}=0$ we note that if $1=\mathbb{E}_{{\cal C},h}[\langle s_o\rangle_{\cal C}]$ then 
$\mathbb{E}_{{\cal C},h}[\langle s_o\rangle_{\cal C}^2] - \mathbb{E}_{{\cal C},h}[\langle s_o\rangle_{\cal C}]^2 = 0$ because of the Nishimori identity
$\mathbb{E}_{{\cal C},h}[\langle s_o\rangle_{\cal C}^2] = \mathbb{E}_{{\cal C},h}[\langle s_o\rangle_{\cal C}]$. Thus the random variable $\langle s_o\rangle_{\cal C}$
does not fluctuate and equals $1$ almost surely. Thus $\sign \langle s_o\rangle_{\cal C}= +1$ and $P_{e}=0$. For the converse we combine Fano's inequality \cite{Ash} together with
Jensen to get $0\leq \frac{1}{N}H({\bf X}\vert {\bf Y})\geq h(P_{error})$ where $h$ is the binary entropy function. 
Thus $\lim_{N\to+\infty}\frac{1}{N} H({\bf X}\vert {\bf Y})=0$. If this is true for a whole range of $n$ we can conclude $g_{MAP}(n)=0$.

Combining \eqref{correlation} and \eqref{identity} we obtain \cite{Macris2}, \cite{Macris1}

\vskip 0.25cm

\noindent{\bf Theorem. }{\it 
Assume communication through a BIAWGNC with noise $n$ and an LDPC ensemble of codes. Then
\begin{equation}\label{main}
g_{MAP}(n)\geq \lim_{d\to+\infty}\frac{1}{2n^3}\mathbb{E}_{ h,l, u_1,...,u_l}
\biggl[1-\tanh(h+\sum_{c=1}^{l}u_c^{(d)})\biggr]
\end{equation}
where the right hand side is computed from the BP 
algorithm and defines the generalized EXIT curve associated to the BP decoder, $g_{BP}(n)$. The p.d.f of $h$ is gaussian with mean and variance $n^{-1/2}$, $l$ is
the random degree of variable nodes, the distribution of $u_i$ is induced by the message passing algorithm.}

\vskip 0.25cm

Such bounds and the method used here extends to the class of (smooth) binary input symmetric channels \cite{Macris3}. These bounds have also been derived 
recently by the method
of physical degradation \cite{Urbankebook}, \cite{MMUR}. To conclude we briefly discuss a number of consequences of the theorem.
 
\noindent{\it General picture.} 
In general the BP and MAP curves may have several discontinuites corresponding to several phase transitions in the spin glass. 
In the simplest (non trivial) case where there
is only one discontinuity their behavior is as follows. For $0<n<n_{BP}$ $g_{BP}(n)=0$, there is a jump discontinuity at $n_{BP}$
and for $n>n_{BP}$ $g_{BP}(n)$ is strictly positive. The same occurs for $g_{MAP}$ but with the jump discontinuity at $n_{MAP}$ and $n_{MAP}>n_{BP}$. 
Moreover the 
$BP$ curve is always under the $MAP$ curve. 

\noindent{\it Bound on MAP threshold.} From the definition of the MAP generalized EXIT curve we see that
\begin{equation}
\int_{n_{MAP}}^{+\infty} g_{MAP}(n) dn = 
\liminf_{N\to\infty}\frac{1}{N}( H({\bf X}\vert {\bf Y})\vert_{n=+\infty} - H({\bf X}\vert {\bf Y})\vert_{n=n_{MAP}})= R
\end{equation}
Indeed for infinite noise we have no knowledge of the sent signal (so the conditional entropy is $R$) and just below the 
MAP threshold we have perfect knowledge (the conditional entropy is zero). The theorem then implies 
\begin{equation}
R<\int_{n_{MAP}}^{+\infty} g_{BP}(n) dn
\end{equation} 
where the rigth hand side can be computed numericaly. This then yields a lower bound on the MAP threshold. Numerical evaluations tend to show that this bound
is tight which suggests that above the MAP threshold the BP and MAP curves should coincide \cite{CyrilRuediger}
(this can be proved for the binary erasure channels
and some codes \cite{MMU2}).

\noindent{\it Bounds on the conditional entropy.} It is possible to obtain bounds on the conditional entropy itself by integration of the 
inequality \eqref{main}. Let us set $h({\bf X}\vert {\bf Y})=\liminf_{N\to+\infty}\frac{1}{N}H({\bf X}\vert {\bf Y})$. Integrating from $0$ to $n$ we get
\begin{equation}\label{previous}
h({\bf X}\vert {\bf Y})\leq \int_0^n  g_{BP}(n) dn
\end{equation}
and integrating from $n$ to $+\infty$,
\begin{equation}\label{last}
h({\bf X}\vert {\bf Y})\geq R - \int_n^{+\infty}  g_{BP}(n) dn= \int_0^n  g_{BP}(n) dn + (R - \int_0^{+\infty} g_{BP}(n) dn)
\end{equation}
In the case where there is no phase transition one can show that $R=\int_0^{+\infty} dn g_{BP}(n)$ so that we get an exact expression
for the conditional entropy and its derivative satisfies $g_{MAP}(n)=g_{BP}(n)$. We have a situation where the model is exactly solved and the result
of the cavity method (or replica symmetric expression) is proved to be exact. However there is no fully polarized phase and no error free communication.
When there is one (or many) phase transition the parenthesis in the last the right hand side of \eqref{last} is strictly negative so that the two bounds 
for $h({\bf X}\vert {\bf Y})$ do not match. However it is believed that the upper bound \eqref{previous} is tight above 
the MAP threshold because it coincides with the result of the replica
symmetric calculation. The same bound has been obtained \cite{Montanari2} using the interpolation methods developped by Guerra \cite{Guerra} for the Sherrington-Kirkpatrick
model. Clearly, it would be desirable to prove the converse inequality.

\end{section}


\begin{thebibliography}{10}

\bibitem{Ash} R. B. Ash, {\it Information Theory} Dover Publications (1990).

\bibitem{Sourlas}
N. Sourlas, Spin glass models as error correcting codes, Nature, {\bf 339} (1989) pp693-695; in From Statistical Physics to Statistical
inference and back, eds. P. Grassberger and J. P. Nadal, Kluwer Academic (1994) p. 195; in Mathematical Results in Statistical
Mechanics, eds S. Miracle-Sole, J. ruiz, V. Zagrebnov, World Scientific (1999) p. 475.

\bibitem{Montanari1} A. Montanari, {\it The glassy phase of Gallager codes}, European Physical Journal, {\bf 23} (2001).

\bibitem{Nishimori} H. Nishimori, {\it Statistical Physics of Spin Glasses and Information Processing: An Introduction}, 
Oxford Science Publications 
(2001).

\bibitem{Urbankebook} R. Urbanke, T. Richardson, in {\it Modern Coding Theory}, Cambridge University Press (in preparation).

 
\bibitem{Guerra} F. Guerra, F. Toninelli, {\it Quadratic replica coupling in the Sherrington-Kirkpatrick 
mean field spin glass model},  J. Math. Phys {\bf 43}, p 3704 (2002)

\bibitem{Montanari2} A. Montanari, {\it Tight bounds for LDPC and LDGM codes under MAP decoding}, preprint(2004)

\bibitem{Contucci} S. Morita, H. Nishimori, P. Contucci, {\it Griffiths inequalities for the Gaussian spin glass}, J. Phys. A 37 (2004) L203.

\bibitem{CyrilRuediger}
C. M\'easson and R. Urbanke,  {\it An upper-bound for the ML threshold of iterative coding systems over the BEC}, 
 Proc. of the 41st Allerton Conference on Communications, Control and Computing, Allerton House, Monticello, USA, 
October 2003  p.3.

\bibitem{Macris1} N. Macris, {\it Correlation inequalities: a useful tool in the theory of LDPC codes}, Proc. IEEE int symp inf theory, Adelaide, September (2005). 

\bibitem{Macris2} N. Macris, {\it Griffith-Kelly-Sherman correlation inequalities: a useful tool in the theory of error correcting codes}, 
preprint (2005)

\bibitem{Macris3} N. Macris, {\it Sharp bounds on generalized EXIT functions}, preprint (2005).

\bibitem{MMUR} C. M\'easson, A. Montanari,and R.Urbanke, Proc of IEEE Int. Symp Inf. Theory, Adelaide, Australia (2005).

\bibitem{CyrilRuediger}
C. M\'easson and R. Urbanke,  {\it An upper-bound for the ML threshold of iterative coding systems over the BEC}, 
 Proc. of the 41st Allerton Conference on Communications, Control and Computing, Allerton House, Monticello, USA, 
October 2003  p.3. 

 
\end{thebibliography}
\end{document}